\documentclass[11pt, a4paper, onecolumn, copyright]{AweAI}

\usepackage[authoryear, sort&compress, round]{natbib}
\usepackage[utf8]{inputenc}
\DeclareUnicodeCharacter{1EAB}{\^a}
\DeclareUnicodeCharacter{1EBF}{\^e}

\usepackage{amsfonts}
\usepackage{amssymb}
\usepackage{graphicx}
\usepackage{amsthm}
\usepackage{subcaption}
\usepackage{multirow}
\usepackage{enumitem}
\usepackage{bbding}
\usepackage{colortbl}
\usepackage{algorithmic}
\usepackage{subcaption}
\usepackage{wrapfig}
\usepackage{tikz}
\usepackage{pgfplots}
\usepackage{bm}
\usepackage[linesnumbered,ruled,vlined]{algorithm2e}
\usepackage{framed,xcolor}
\usepackage[most]{tcolorbox}
\usepackage{booktabs}
\usepackage{adjustbox}
\usepackage{xspace}
\usepackage{float}
\usepgfplotslibrary{groupplots}
\pgfplotsset{compat=1.18}

\newcommand{\indicator}[1]{\mathds{1}}
\newcommand{\paratitle}[1]{\vspace{1.5ex}\noindent\textbf{#1}}

\newcommand{\wo}{\emph{w/o}\xspace}
\newcommand{\w}{\emph{w/}\xspace}

\definecolor{bg_thought}{RGB}{240, 248, 255}
\definecolor{bg_action}{RGB}{255, 245, 238}
\definecolor{bg_obs}{RGB}{245, 245, 245}
\definecolor{bg_io}{RGB}{245, 240, 255}
\definecolor{line_gray}{RGB}{180, 180, 180}
\definecolor{bg_omit}{RGB}{250, 250, 240}
\definecolor{headergray}{RGB}{242, 244, 247}
\definecolor{sectionblue}{RGB}{235, 240, 245}

\title{Learning to Retrieve from Agent Trajectories}
\renewcommand{\today}{}

\newcommand{\githubrepo}{https://github.com/Yuqi-Zhou/LRAT}
\newcommand{\homepageurl}{https://yuqi-zhou.github.io/LRAT-homepage/}
\newcommand{\hfcollectionurl}{https://huggingface.co/collections/Yuqi-Zhou/lrat}
\makeatletter
\renewcommand{\abscontent}{%
    \noindent
    \parbox{\dimexpr\linewidth}{\absfont \theabstract}%
    \vskip1em
    \noindent  GitHub: \href{\githubrepo}{\texttt{https://github.com/Yuqi-Zhou/LRAT}}\\
    \noindent  Homepage: \href{\homepageurl}{\texttt{https://yuqi-zhou.github.io/LRAT-homepage/}}\\
    \noindent  Collection: \href{\hfcollectionurl}{\texttt{https://huggingface.co/collections/Yuqi-Zhou/lrat}}%
}
\fancypagestyle{firststyle}{
    \fancyhead[L]{%
        \raisebox{-0.5\height}{%
            \includegraphics[
                height=20pt,
                keepaspectratio
            ]{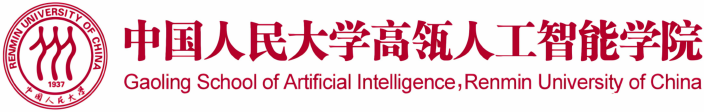}%
        }%
    }

    \fancyhead[R]{%
        \ifdefined\paperurl
        \if\relax\the\paperurl\relax \else
            \href{\the\paperurl}{\urlheaderfont \itshape \the\paperurl}\\ \fi
        \else
        \fi
        {\footerfont\itshape\monthyeardate\today\hspace{8pt}}%
        \raisebox{-0.5\height}{%
            \includegraphics[
                height=23pt,
                keepaspectratio
            ]{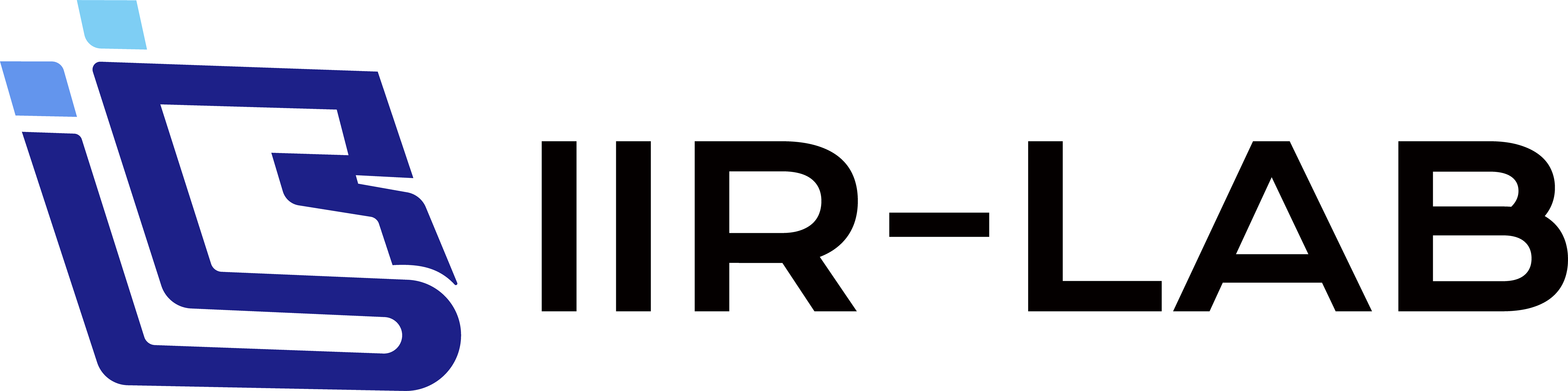}%
        }%
        \ifthenelse{\boolean{confidential}}{\\ \footerfont \internalonly}{}
    }

    \fancyhead[C]{}

    \fancyfoot[L]{
    \footerfont\itshape
        \ifdefined\correspondingauthor
        \if\relax\detokenize\expandafter{\the\correspondingauthor}\relax
      \else
        \footerfont\itshape {Emails: \the\correspondingauthor\\}%
      \fi
    \fi
    }

    \fancyfoot[R]{
        \ifthenelse{\boolean{confidential}}{
        \ifdefined\reportnumber
        \if\relax\the\reportnumber\relax
        \else \footerfont\itshape  {\footerfont \thepa{} Technical Report \the\reportnumber} \fi
        \else \fi
        }{\footerfont\bfseries\relax}
    }
    \fancyfoot[C]{
        \footerfont\bfseries\relax}
}
\makeatother

\author[1]{Yuqi Zhou\textsuperscript{*}}
\author[1]{Sunhao Dai\textsuperscript{*\dag}}
\author[1]{Changle Qu}
\author[2]{Liang Pang}
\author[1]{Jun Xu\textsuperscript{\Envelope}}
\author[1]{Ji-Rong Wen}
\correspondingauthor{yuqizhou@ruc.edu.cn, sunhaodai@ruc.edu.cn, junxu@ruc.edu.cn}
\affil[1]{Gaoling School of Artificial Intelligence, Renmin University of China}
\affil[2]{CAS Key Laboratory of AI Safety, Institute of Computing Technology, Chinese Academy of Sciences\footnote{\textsuperscript{*}Equal contribution. \textsuperscript{\dag}Project Leader. \textsuperscript{\Envelope}Corresponding author.}}

\begin{abstract}
Information retrieval (IR) systems have traditionally been designed and trained for human users, with learning-to-rank methods relying heavily on large-scale human interaction logs such as clicks and dwell time. With the rapid emergence of large language model (LLM) powered search agents, however, retrieval is increasingly consumed by agents rather than human beings, and is embedded as a core component within multi-turn reasoning and action loops. In this setting, retrieval models trained under human-centric assumptions exhibit a fundamental mismatch with the way agents issue queries and consume results.
In this work, we argue that retrieval models for agentic search should be trained directly from agent interaction data. We introduce \emph{learning to retrieve from agent trajectories} as a new training paradigm, where supervision is derived from multi-step agent interactions. Through a systematic analysis of search agent trajectories, we identify key behavioral signals that reveal document utility, including browsing actions, unbrowsed rejections, and post-browse reasoning traces.
Guided by these insights, we propose LRAT, a simple yet effective framework that mines high-quality retrieval supervision from agent trajectories and incorporates relevance intensity through weighted optimization.
Extensive experiments on both in-domain and out-of-domain deep research benchmarks demonstrate that retrievers trained with LRAT consistently improve evidence recall, end-to-end task success, and execution efficiency across diverse agent architectures and scales. Our results highlight agent trajectories as a practical and scalable supervision source, pointing to a promising direction for retrieval in the era of agentic search.
\end{abstract}

\begin{document}

\begingroup
\makeatletter
\renewcommand{\thefootnote}{}
\renewcommand{\@makefnmark}{}
\maketitle
\makeatother
\endgroup

\begin{figure}[h]
    \centering
    \includegraphics[width=\linewidth]{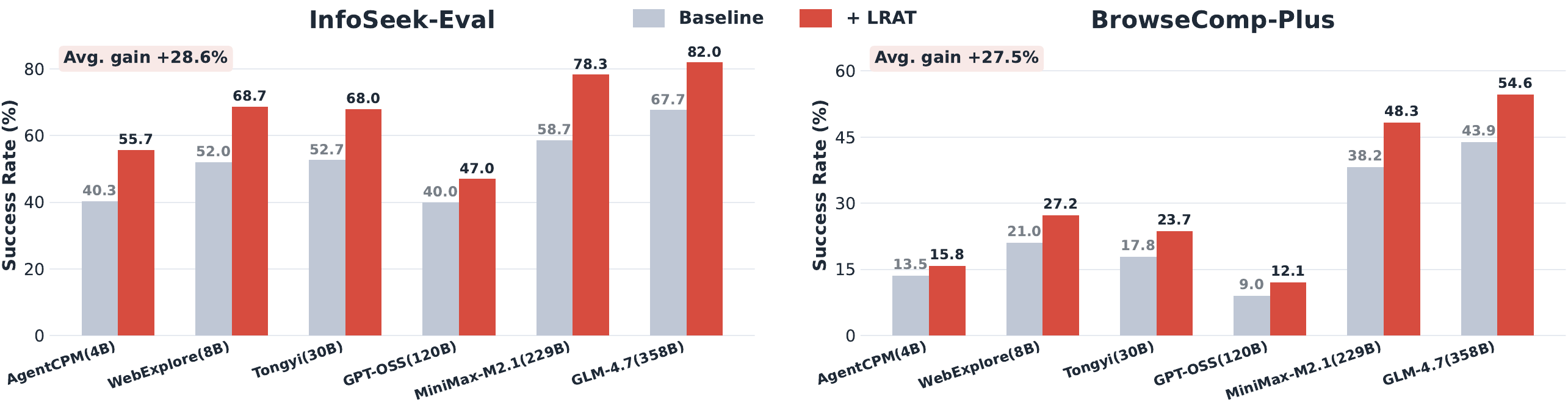}
    \vspace{-0.4em}
    \caption{
    Overview of LRAT gains across six agent backbones with Qwen3-Embedding-0.6B as the retriever.
    Left: Success rate on InfoSeek-Eval. 
    Right: Evidence recall on BrowseComp-Plus. 
    LRAT consistently improves both in-domain task success and out-of-domain retrieval quality.
    }
    \label{fig:hero_intro}
    \vspace{-0.4em}
\end{figure}

\section{Introduction}

\begin{wrapfigure}{r}{0.52\textwidth}
    \vspace{-0.5em}
    \centering
    \includegraphics[width=\linewidth]{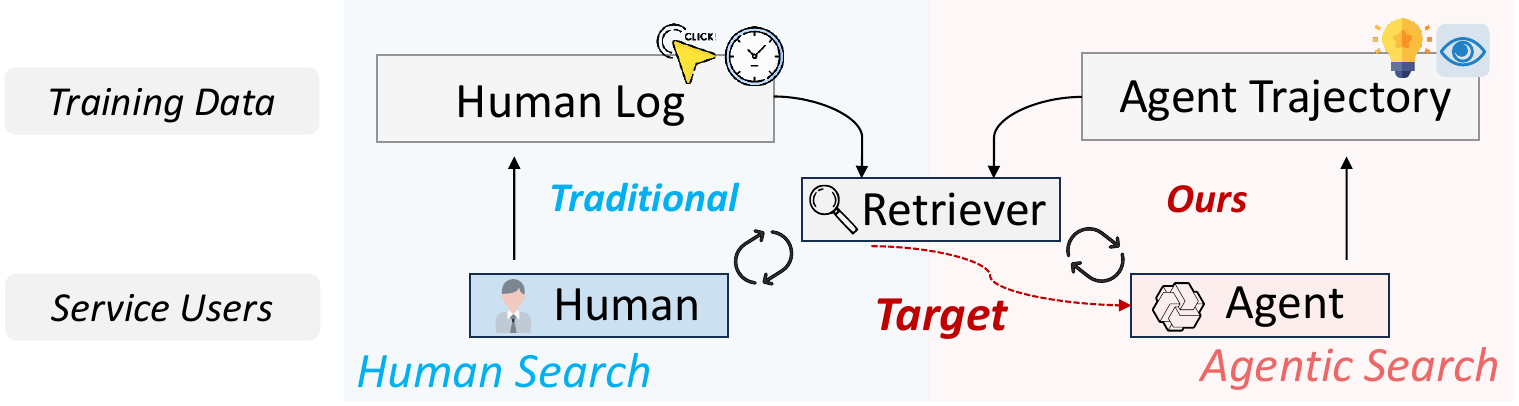}
    \vspace{-0.8em}
    \caption{Illustration of the search paradigm shift where the retriever’s target changes from humans to agents.}
    \label{fig:intro}
    \vspace{-0.8em}
\end{wrapfigure}
    
Information retrieval (IR) has long served as the foundation of information access systems such as web search engines~\cite{baeza1999modern, manning2009introduction, chowdhury2010introduction, croft2010search}, and decades of research in learning to retrieve and rank have been built around a human-centric paradigm~\cite{xu2007adarank,cao2006adapting,liu2009learning}. 
In this setting, retrieval models are trained from large-scale human interaction logs (e.g., clicks~\cite{joachims2002optimizing, joachims2017accurately} and dwell time~\cite{kelly2004display, kim2014modeling}) and optimized to serve humans~\cite{joachims2002optimizing, joachims2017accurately, shen2005implicit}, forming a powerful data flywheel.
With the rapid emergence of large language model (LLM) powered agents, however, this paradigm is being fundamentally challenged~\cite{ shi2025deep,xu2025comprehensive,huang2025deep}. 
Search engines are increasingly queried by agents rather than humans, and retrieval is no longer a standalone endpoint but a core tool embedded within an agent's multi-turn reasoning and action loop~\cite{dai2025next, xi2025survey, white2024advancing}. 
Search agents iteratively issue sub-queries, consume retrieved information, and refine their actions to solve complex tasks, making retrieval quality a critical bottleneck that directly constrains what information agents can observe, reason over, and ultimately accomplish.

Despite this shift in usage, today’s search agents typically rely on general-purpose retrieval models~\cite{jin2025searchr,song2025r1,wang2025information}, such as dense embedding retrievers, or external search APIs (e.g., Google or Bing). 
As shown in \autoref{fig:intro}, these retrievers are overwhelmingly trained from human interaction logs and implicitly encode human-centric assumptions about how queries are issued, how results are examined, and how relevance signals are expressed. 
When the primary user becomes an agent, these assumptions break in fundamental ways.
Agent queries are not issued to satisfy immediate informational needs, but to advance intermediate reasoning objectives during problem-solving, resulting in relevance patterns that differ from those of human users.
As a result, search agents are still served by retrieval models that are trained primarily from human data, creating a fundamental mismatch between how retrieval models are trained and how they are used in agentic search.

This mismatch motivates a rethinking of retrieval training for the agent era.
In this work, we argue that retrieval models should be trained directly from agent interaction data, rather than repurposed from human-centric search.
Analogous to how learning to retrieve and rank from user interaction logs in web search, we propose \textbf{learning to retrieve from agent trajectories} as a new training paradigm.
Agent trajectories record the sequence of intermediate queries, retrieved documents, and reasoning steps generated during task execution, providing a rich and naturally abundant source of supervision.
Training retrievers on such trajectories intrinsically aligns retrieval objectives with agents’ feedback.
Moreover, because agent trajectories are generated as a byproduct of every agent invocation, they create an opportunity to build a sustainable data flywheel for retrieval in agentic search, positioning trajectories as the agent-era counterpart of user click logs.

In this paper, we take an early step toward this direction by systematically analyzing how deep research agents interact with retrieval systems through multi-turn execution trajectories. 
Our analysis reveals fundamental differences between human feedback and agent feedback, and identifies key agent behaviors that reliably indicate document utility. 
Specifically, we show that (1) browsing actions constitute a necessary condition for task success, making browsed documents natural candidates for naive positive signals; (2) unbrowsed documents in agent trajectories serve as reliable negatives without position bias; and (3) post-browse reasoning traces provide a strong indicator of relevance intensity, distinguishing truly useful documents from superficially browsed ones. Guided by these insights, we propose \textbf{LRAT} (\textbf{L}earning to \textbf{R}etrieve from \textbf{A}gent \textbf{T}rajectories), a simple yet effective framework for training retrieval models directly from agent interactions. 
LRAT first mines coarse query-document supervision from search-browse transitions and refines positive signals using explicit post-browse reasoning traces. 
It then incorporates relevance intensity through reasoning-length–aware weighting to prioritize documents that drive substantial agent progress.
In our instantiation, LRAT is applied to trajectories generated by Tongyi-DeepResearch-30B on 10K InfoSeekQA queries with four retrievers, yielding \textbf{26,482} agent trajectories and \textbf{91,713} training pairs.
Importantly, LRAT requires no additional human annotation and can be applied to trajectories generated by arbitrary agents and retrievers, making it a practical and scalable approach for agent-centric retrieval training.

In summary, our main contributions are threefold:

$\bullet$ 
We identify a fundamental misalignment between human-centric retrieval training and agentic search, and formulate \emph{learning to retrieve from agent trajectories} as a new retrieval paradigm. In this setting, supervision is derived from multi-step agent interactions, reflecting how retrieval is actually used by search agents.

$\bullet$ 
Guided by insights from empirical analysis, we propose LRAT, a simple yet effective framework that mines high-quality retrieval supervision from agent trajectories, providing a practical step toward agent-aligned retriever training.

$\bullet$  
Experiments on both in-domain and out-of-domain deep research benchmarks show that LRAT consistently improves evidence retrieval and end-to-end agent performance across diverse agent architectures and scales. We further demonstrate that LRAT can support a self-improving data flywheel, highlighting the scalability value of LRAT in real-world scenarios.
\section{Related Work}

\paratitle{Search Agent \& Retriever Optimization for RAG.}
The rise of agentic search, exemplified by deep search agents such as Search-o1~\cite{li-etal-2025-search} and Search-R1~\cite{jin2025searchr}, has revolutionized how LLMs handle complex information-seeking tasks through multi-step reasoning and iterative interactions with search engines.
While significant progress has been made in optimizing the agent~\cite{zheng2025deep,team2025tongyi,wang2025information}, the underlying retrieval model has largely been treated as a static, off-the-shelf tool (e.g., Google Search API, BM25, or dense embedding models).
This overlooks a fundamental mismatch between static retrievers and the interactive, multi-step nature of agentic search, causing retrieval to emerge as the primary performance bottleneck for search agents.
Although traditional Retrieval-Augmented Generation (RAG) research has explored retriever optimization through preference alignment~\cite{qu-etal-2025-uplift,shi-etal-2024-replug,dong2025understand}, these methods are primarily designed for single-turn retrieval based on a user's initial, explicit query.
In contrast, agentic search operates in a multi-step and interactive manner. 
Search queries are not the fixed initial user query but intermediate actions generated by the agent during reasoning.
These queries are highly dynamic and cannot be judged as correct based on a final answer, unlike prior work~\cite{salemi2024evaluating,zamani2024stochastic,ke-etal-2024-bridging,zhang2024large}. In contrast to prior work, our work shifts the focus from merely refining the agent to optimizing the retriever, enabling it to respond effectively to the agents’ information needs as they evolve across multiple turns.

\paratitle{Learning to Retrieve/Rank from Human Feedback.}
Learning to retrieve/rank has a long history in the information retrieval community~\cite{xu2007adarank,cao2006adapting,liu2009learning,cao2007learning}.
Early work relies on explicit relevance judgments annotated by human assessors, forming the foundation of supervised learning-to-rank methods that optimize document ordering using pointwise, pairwise, or listwise objectives~\cite{burges2005learning,li2007mcrank,burges2010ranknet}.
To overcome the scalability bottlenecks of manual annotation, a significant body of research shifted toward implicit user feedback, leveraging interaction signals such as clicks \cite{joachims2002optimizing, joachims2017accurately}, dwell time \cite{kelly2004display, kim2014modeling}, and scrolling patterns \cite{agichtein2006learning,agichtein2006improving,radlinski2008does}. 
These methods treat user interaction logs as weak supervision signals, employing click models or counterfactual learning to refine ranking models at scale.
More recently, the primary users of modern search systems are no longer limited to humans, but increasingly include autonomous agents that interact with search engines.
Motivated by this shift, our work investigates learning to retrieve from agent trajectories, where supervision is derived from the agent’s sequential interactions with the search system.
\section{Preliminaries}\label{sec:preliminary}

\subsection{Deep Research Agent Trajectories} \label{sec:agent_trajectories}

In this paper, we focus on \emph{Deep Research Agents}, i.e., LLM-powered intelligent agents that solve complex information-seeking tasks via iterative interactions with an external retrieval system. A typical execution trajectory of a Deep Research Agent is illustrated in \autoref{fig:trajectory_example}.

Formally, given an initial user query $q$, a Deep Research Agent follows a ReAct-style~\cite{yao2022react} interaction pattern, repeatedly alternating between \emph{thinking} and \emph{acting}, which yields a multi-turn execution trajectory
$\mathcal{T} = \{(r_t, a_t, o_t)\}_{t=1}^{T}$.
At each turn $t$, the agent maintains an internal reasoning state $r_t$ that guides subsequent actions $a_t$, incorporates new observed information $o_t$ returned from the external retrieval system, and determines whether further information is required.

\textbf{[Think]}: At the beginning of each turn $t$, the agent first produces an explicit reasoning state (or \emph{thought}) $r_t$. This reasoning step analyzes the current context, summarizes previously acquired evidence, and identifies what information is still missing for answering the original user query.

\begin{figure}[t]
    \centering
    \begin{tcolorbox}[colback=white, colframe=gray!60!black, 
                      title=Trajectory Example, 
                      fonttitle=\bfseries, fontupper=\footnotesize,
                      arc=1mm, boxrule=0.8pt, 
                      left=0.4mm, right=0.4mm, top=1mm, bottom=1mm]

    \noindent\colorbox{bg_io}
    {\parbox{\dimexpr\linewidth-2\fboxsep}{\textbf{{User Query:}} Which station merged with Immaculate Heart Radio in 2017?}}
    \vspace{-3pt}\hrule{\color{line_gray}}\vspace{1pt}

    \noindent\colorbox{bg_omit!10}{
            \parbox{\dimexpr\linewidth-2\fboxsep}{
                \centering \scriptsize \textcolor{black!80!black}{
                    \textbf{Agent Trajectory}
                }
            }
        }
    \vspace{-4pt}\hrule{\color{line_gray}}\vspace{1pt}

    \noindent\colorbox{bg_thought}{\parbox{\dimexpr\linewidth-2\fboxsep}{\textbf{{[Think]:}} No merger info yet. Let's search for the 2017 merger details.}}
    
    \noindent\colorbox{bg_action}{\parbox{\dimexpr\linewidth-2\fboxsep}{\textbf{[Search]:} \texttt{Search["Immaculate Heart" "merged" "2017"]}}}
    
    \noindent\colorbox{bg_obs}{\parbox{\dimexpr\linewidth-2\fboxsep}{\textbf{[Information]:} Found 10 results:\\
    \textbf{Doc~[3948]:} Immaculate Heart Radio ... \\
    \textbf{Doc~[2295]:} corporate name Relevant Radio, Inc... \\
    \textbf{...}
    } \\
    }
    
    \vspace{2pt}\hrule{\color{line_gray}}\vspace{2pt}

    \noindent\colorbox{bg_thought}{\parbox{\dimexpr\linewidth-2\fboxsep}{\textbf{[Think]:} Doc~[2295] mentioned Relevant Radio. It might contain the specific merger date. Let's read it.}}
    
    \noindent\colorbox{bg_action}{\parbox{\dimexpr\linewidth-2\fboxsep}{\textbf{[Browse]:} \texttt{Browse[DocID: 2295]}}}
    
    \noindent\colorbox{bg_obs}{\parbox{\dimexpr\linewidth-2\fboxsep}{\textbf{[Information]:} ... Relevant Radio announced it would merge...}}

    \vspace{1pt}\hrule{\color{line_gray}}\vspace{1pt}

    \noindent\colorbox{bg_omit!10}{
            \parbox{\dimexpr\linewidth-2\fboxsep}{
                \centering \scriptsize \textcolor{gray!80!black}{
                    \textit{... Intermediate turns  omitted ...}
                }
            }
        }
    \vspace{-4pt}\hrule{\color{line_gray}}\vspace{1pt}

    \noindent\colorbox{bg_thought}{\parbox{\dimexpr\linewidth-2\fboxsep}{\textbf{{[Think]:}} I have found the answer...}}

    \noindent\colorbox{bg_io}{\parbox{\dimexpr\linewidth-2\fboxsep}{\textbf{[Answer]:} The station is \textcolor{red}{WMJR} (Nicholasville, KY).}}

    \end{tcolorbox}
    
    \vspace{-8pt}
    \caption{An example of Deep Research Agent trajectory.}
    \vspace{-0.4cm}
    \label{fig:trajectory_example}
\end{figure}

\textbf{[Search] or [Browse]}: Conditioned on this reasoning state, the agent selects an action $a_t \in \mathcal{A}$ to interact with the external retrieval environment. We consider two typical actions:

$\bullet$ When invoking the \textbf{[Search]} action, the agent generates an intermediate search query $q_t$ that targets a specific information gap identified in the reasoning state $r_t$. In response, the retrieval system returns a ranked set of top-$K$ candidate documents $\mathcal{D}_t = \{ d_{t,i} \}_{i=1}^{K}$.
In practice, the agent does not immediately observe the full content of each document. Instead, it receives \textbf{a snippet list} for $\mathcal{D}_t$ (e.g., titles and brief summaries) as observation $o_t$, which provides coarse-grained signals for initial assessment and subsequent decision making.

$\bullet$ When invoking the \textbf{[Browse]} action, the agent selects one document $d_t$ from the previously retrieved candidates $\mathcal{D}_{t’} (t' < t)$ and requests to read it in full. The retrieval system then returns the \textbf{complete content} of the selected document as the observed information $o_t$ for this turn. The agent incorporates this information into its reasoning state and continues the cycle of reasoning and action across multiple turns.

\textbf{[Answer]}: This iterative process continues until the agent determines that sufficient information has been gathered to answer the original query.  
At termination, the agent performs a final reasoning step that synthesizes the accumulated evidence across the trajectory and generates the final answer, which is returned to the user.

\subsection{Task Definition} \label{sec:task_definition}

Most existing search agents employ off-the-shelf retrieval models, such as pretrained dense retrievers (e.g., Qwen3-Embedding or E5-Embedding) or external search APIs (e.g., Google or Bing). These retrievers are typically trained from human interaction logs and optimized for human-centric search behavior. In contrast, deep research agents generate rich multi-turn execution trajectories that reveal how retrieval results are examined and utilized. Despite being naturally abundant, such trajectory data is largely untapped for training retrieval models. This raises a natural question: \emph{can retriever be trained directly from agent trajectories?}

We define the task of \emph{Learning to Retrieve from Agent Trajectories} as follows: given a collection of agent execution trajectories $\{\mathcal{T}\}$ as defined in Section~\ref{sec:agent_trajectories}, the goal is to learn a retrieval model that produces ranked document lists optimized to support an agent’s multi-step reasoning and problem-solving process. Unlike traditional learning-to-rank, supervision in this task should be directly derived from agent trajectories, aligning retrieval training directly with agent behaviors.

\section{Analysis of Agent Trajectories}\label{sec:analysis}

In this section, we conduct a systematic analysis of deep research agent trajectories.  
Our primary goal is to understand how agents interact with retrieval systems in practice, identify patterns that differ from human search, and extract insights that inform the design of agent-centric relevance modeling.  

\subsection{Environment Setup}\label{sec:env_setup}
This section describes the experimental setup used to generate and analyze deep research agent trajectories.

\begin{table*}[t]
    \centering
    \small
    \caption{
    Statistics of generated trajectories across different retrievers.
    For each retriever, we report the number of completed trajectories ($N$) and the average numbers of [Search] actions ($\text{Avg.\ }S$), [Browse] actions ($\text{Avg.\ }B$), their ratio ($B/S$), and the total execution steps ($\text{Avg.\ }T$),
    categorized into correct, incorrect, and all completed trajectories.
    }
    \vspace{-0.2cm}
    \label{tab:trajectory_statistics}
    \setlength{\tabcolsep}{2.2pt}
    \resizebox{\textwidth}{!}{
    \begin{tabular}{l ccccc ccccc ccccc}
        \toprule
        \multirow{2}{*}{\textbf{Retriever}} 
        & \multicolumn{5}{c}{\textbf{Correct}} 
        & \multicolumn{5}{c}{\textbf{Incorrect}} 
        & \multicolumn{5}{c}{\textbf{Total (Complete)}} \\
        \cmidrule(lr){2-6} \cmidrule(lr){7-11} \cmidrule(lr){12-16}
        & \textbf{$N$} & \textbf{Avg.\ $S$} & \textbf{Avg.\ $B$} & \textbf{$B$/$S$} & \textbf{Avg.\ $T$}
        & \textbf{$N$} & \textbf{Avg.\ $S$} & \textbf{Avg.\ $B$} & \textbf{$B$/$S$} & \textbf{Avg.\ $T$}
        & \textbf{$N$} & \textbf{Avg.\ $S$} & \textbf{Avg.\ $B$} & \textbf{$B$/$S$} & \textbf{Avg.\ $T$} \\
        \midrule
        BM25        & 7,674 & 9.15 & 2.96 & 0.32 & 12.11 & 1,872 & 29.15 & 5.97 & 0.20 & 35.11 & 9,546 & 13.07 & 3.55 & 0.27 & 16.63 \\
        Qwen3-Embedding-0.6B  & 5,913 & 12.81 & 3.68 & 0.29 & 16.49 & 2,062 & 38.95 & 7.17 & 0.18 & 46.12 & 7,975 & 19.57 & 4.58 & 0.23 & 24.15 \\
        Qwen3-Embedding-4B    & 6,354 & 13.24 & 4.11 & 0.31 & 17.34 & 2,121 & 36.13 & 7.47 & 0.21 & 43.60 & 8,475 & 18.97 & 4.95 & 0.26 & 23.91 \\
        Qwen3-Embedding-8B    & 6,541 & 11.86 & 3.69 & 0.31 & 15.55 & 2,082 & 34.47 & 7.20 & 0.21 & 41.67 & 8,623 & 17.32 & 4.54 & 0.26 & 21.85 \\
        \midrule
        \textbf{Total} 
        & \textbf{26,482} & 11.77 & 3.61 & 0.31 & 15.38 
        & \textbf{8,137}  & 34.68 & 6.95 & 0.20 & 41.63 
        & \textbf{34,619} & 17.25 & 4.41 & 0.26 & 21.66 \\
        \bottomrule
    \end{tabular}
    }
    \vspace{-0.1cm}
\end{table*}

\subsubsection{Seed Data Selection} \label{sec:env_data}
To encourage sustained search and browsing behavior, we adopt InfoSeekQA~\cite{xia2025open} as the seed dataset for trajectory generation. InfoSeekQA is a large-scale deep research benchmark comprising over 50K question–answer pairs, designed to require hierarchical reasoning and iterative information acquisition. Tasks in InfoSeekQA typically involve substantially deeper search processes, resulting in significantly longer interaction trajectories than traditional QA datasets.  
Given computational constraints and evaluation reliability, we select the top 10K queries with verified ground-truth answers as the seed set for trajectory construction.
Following the standard practice, we use the Wiki-25-Dump\footnote{\url{https://huggingface.co/datasets/Lk123/wiki-25-512}} as the corpus, which comprises over 11.2 million document chunks, each truncated to 512 tokens.

\subsubsection{Retrieval Systems}   
To cover a diverse range of retrieval behaviors, we deploy four retrieval models: a sparse \textit{BM25} retriever~\cite{robertson2009probabilistic} and three dense retrievers of increasing capacity, \textit{Qwen3-Embedding-0.6B, 4B, and 8B}~\cite{yang2025qwen3}, spanning lexical matching and semantic retrieval capabilities.

During trajectory generation, each [Search] action returns the top-$10$ candidate documents. For each candidate, the agent observes a short snippet consisting of the first 64 tokens of the document, which approximates the average length of web search snippets measured under the Qwen3 tokenizer. This design simulates a realistic search environment where agents initially rely on coarse-grained evidence before deciding whether to browse full documents.

\subsubsection{Deep Research Agent Configuration}
We adopt \textit{Tongyi-DeepResearch-30B-A3B}\footnote{\url{https://modelscope.cn/models/iic/Tongyi-DeepResearch-30B-A3B}}~\cite{team2025tongyi} as the search agent for trajectory collection, which is one of the strongest open-source search agents tailored for long-horizon, deep information-seeking tasks and supports over one hundred interaction steps. 
To allow sufficient exploration, we set the maximum number of interaction rounds to $T=100$. If the agent fails to reach a conclusion within this limit, it is required to produce a final answer based on the information collected so far. A trajectory is considered valid only if the final answer matches the ground truth provided by InfoSeekQA.

\subsubsection{Trajectory Generation} 
Using the environment described above, we generate agent trajectories by executing the search agent on each seed query. For each question, the agent iteratively produces reasoning traces and executes [Search] or [Browse] actions until termination. \autoref{fig:trajectory_example} illustrates a representative execution trajectory.  
We filter out trajectories that exceed the maximum step limit or produce incorrect final answers. Answer correctness is verified by comparing the agent’s output against the ground truth using \textit{Qwen3-30B-A3B-Thinking-2507}\footnote{\url{https://huggingface.co/Qwen/Qwen3-30B-A3B-Thinking-2507}}~\cite{yang2025qwen3}.  
The resulting trajectories form the basis of our analysis, and their statistics are summarized in \autoref{tab:trajectory_statistics}.

\subsection{Agent Trajectory Analysis} \label{sec:traj_analysis}

\begin{figure*}[t]
    \centering
    \includegraphics[width=\linewidth]{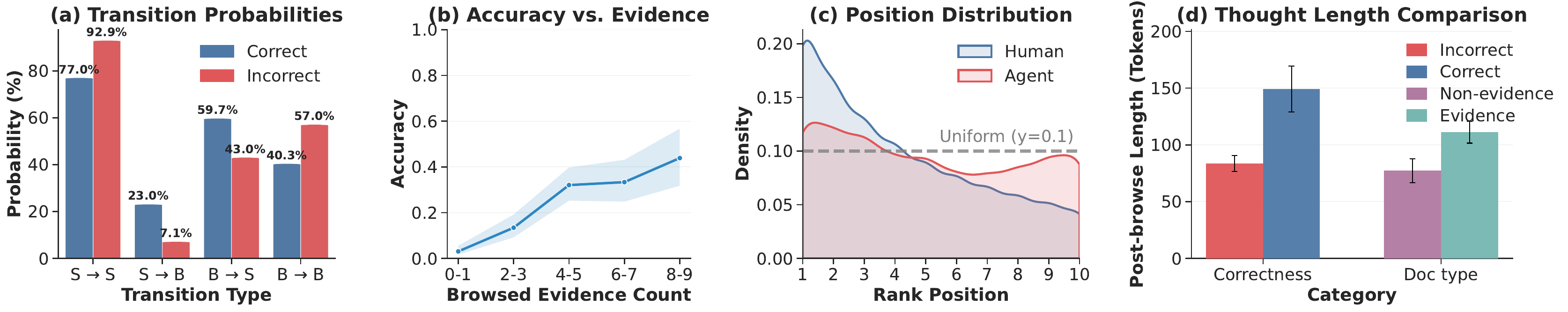}
    \vspace{-15pt}
    \caption{Trajectory analysis on BrowseComp-Plus, where each user query is annotated with supporting evidence documents.
Interaction traces are generated by \textit{Tongyi-DeepResearcher} with the \textit{Qwen3-Embedding-0.6B} retriever.
Across all panels, \textit{correct} and \textit{incorrect} denote trajectories leading to correct and incorrect final answers.
(a) Action transition probabilities between [Search] (S) and [Browse] (B).
(b) Accuracy versus the number of unique evidence documents browsed (binned; shaded areas indicate 95\% Wilson confidence intervals).
(c) Browsing rank-position distribution under shuffled documents (agent vs.\ human \cite{craswell2008experimental}).
(d) Length of post-browse reasoning (tokens) by trajectory correctness (correct vs.\ incorrect) and document type (evidence vs.\ non-evidence).}
    \label{fig:traj_analysis}
    \vspace{-0.1cm}
\end{figure*}

In this section, we analyze deep research agent trajectories to understand how agents interact with retrieval systems during long-horizon problem solving. 

\subsubsection{Browsing Is a Necessary Signal for Successful Retrieval}
\label{sec:browse_signal}

We first investigate which agent behaviors are most indicative of effective information acquisition.  
By comparing successful and failed trajectories, we observe a clear behavioral divergence in how agents interact with the retrieval system.  
As summarized in \autoref{tab:trajectory_statistics}, failed (i.e., incorrect) trajectories exhibit a substantially lower ratio between [Browse] and [Search] actions (B/S).  
This pattern indicates that agents in unsuccessful runs frequently issue search queries but rarely proceed to browse retrieved documents, suggesting that the returned snippets are often deemed insufficiently informative to warrant further inspection.  
In contrast, successful (i.e., correct) trajectories show fewer repetitive search actions and a markedly higher frequency of browsing behaviors, reflecting more effective utilization of retrieved results.

This pattern is further confirmed by the action transition statistics shown in \autoref{fig:traj_analysis}(a). Successful trajectories show a significantly higher probability of transitioning from [Search] (S) to [Browse] (B), whereas unsuccessful trajectories are more likely to remain in search-only loops without progressing to document consumption.  
Moreover, \autoref{fig:traj_analysis}(b) shows that task success increases monotonically with the number of browsed evidence documents, and drops to zero when the agent never browses any document containing the required evidence. 

Together, these results indicate that browsing retrieved documents is not merely correlated with task success, but constitutes a necessary condition for successful task completion. This observation motivates \textbf{treating browsed documents as primary candidates for positive supervision} when training retrieval models for search agents.

\subsubsection{Unbrowsed Documents Are Reliable Negatives} \label{sec:negative_signal}
Having identified browsing as a strong indicator of positive utility, we next examine whether documents that are not browsed can be interpreted as negative signals in agent trajectories. In human-centric click logs, negative signals are notoriously ambiguous due to well-known position bias or exposure bias~\cite{joachims2017accurately}: unclicked documents may be irrelevant or simply unseen. As a result, learning-to-rank methods typically adopt conservative heuristics, such as ``skip-above'' sampling~\cite{joachims2002optimizing, joachims2017accurately,granka2004eye}, to avoid introducing false negatives.

To assess whether similar issues arise in agent trajectories, we analyze the distribution of browsing actions across ranking positions. \autoref{fig:traj_analysis}(c) shows that, unlike human clicks, the agent’s browsing behavior is not sharply concentrated at top ranks. Instead, browsing is distributed relatively evenly across positions, indicating that the agent actively evaluates candidates beyond the highest-ranked results rather than relying on positional cues. This suggests that unbrowsed documents are typically the result of explicit rejection after inspection, rather than limited exposure. Therefore, in contrast to human click logs, \textbf{all unbrowsed documents within a retrieved candidate set can be treated as reliable negatives} without requiring position bias correction.

\subsubsection{Post-Browse Reasoning Traces Are Important Indicators} \label{sec:reasoning_signal}
While browsing behavior identifies documents that the agent chooses to inspect, it remains an implicit signal and may include noise.  
Deep research agents, however, generate explicit reasoning traces immediately following a [Browse] action, providing a more direct view of how retrieved content is interpreted and utilized during problem solving.

We analyze the agent’s post-browse reasoning to understand their relationship with task success and document utility.  
As shown in \autoref{fig:traj_analysis}(d), trajectories that ultimately produce correct answers are associated with significantly longer reasoning following browsing actions than those that lead to incorrect answers.  
A closer inspection of cases reveals a consistent pattern: in unsuccessful trajectories, the agent often quickly abandons a browsed document after determining that it does not contain useful information, resulting in short post-browse reasoning.  
In contrast, successful trajectories exhibit substantially longer reasoning after browsing, reflecting deeper analysis and integration of the retrieved content into subsequent decision making.
Moreover, documents that contain ground-truth evidence are followed by markedly longer reasoning traces than non-evidence documents, indicating that useful content elicits more extensive agent reasoning.  

Together, these observations suggest that \textbf{post-browse reasoning traces provide a reliable signal of document utility}.  
In particular, the length of the reasoning trace after browsing is strongly correlated with whether a document contributes meaningfully to task progress, offering valuable insight into more fine-grained relevance beyond binary feedback.
\section{Learning to Retrieve from Trajectories}\label{sec:method}

\begin{figure*}[t]  
    \centering    
    \includegraphics[width=1\linewidth]{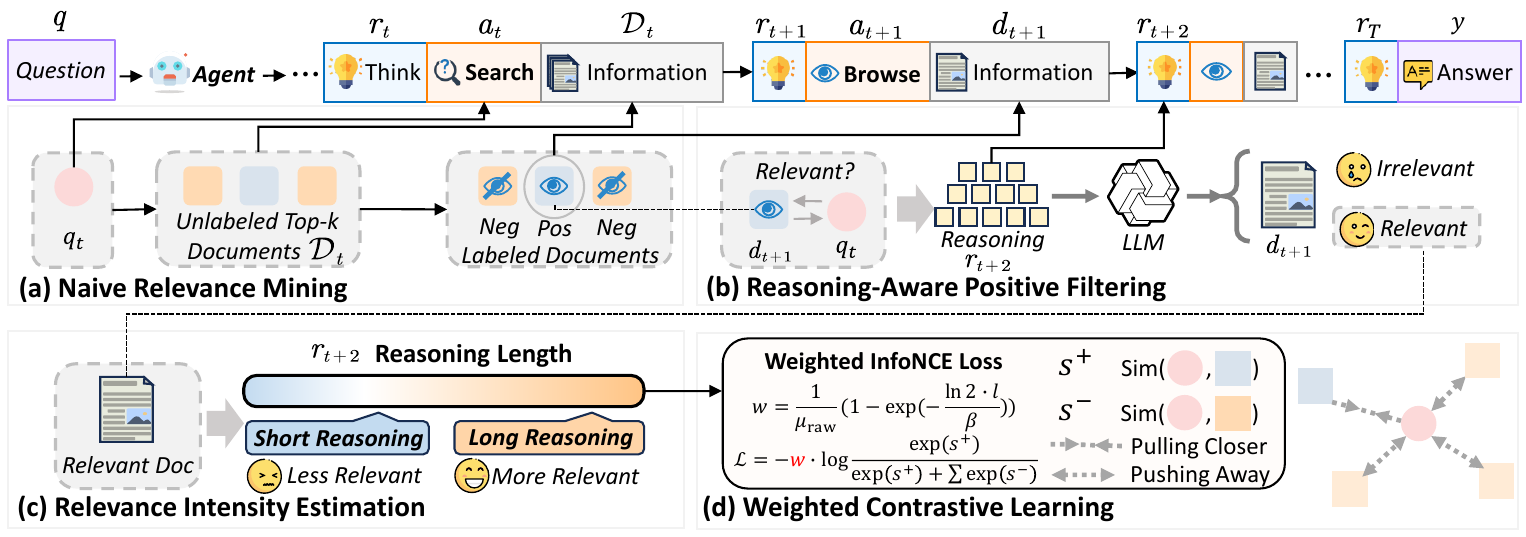}
    \vspace{-18pt}
    \caption{Our proposed LRAT retrievers training framework using deep research agent interactions. (a) Extract relevance signals from \textit{Browse} actions in the interaction sequence. (b) Filter out irrelevant documents by judging \textit{post-Browse reasoning} with LLM. (c) Estimate relevance weights based on the length of \textit{post-Browse reasoning}. (d) Perform contrastive learning using the filtered samples and their relevance weights.}
    \vspace{-10pt}
    \label{fig:method}
\end{figure*}

Motivated by the insights in Section~\ref{sec:analysis}, we propose \textbf{LRAT}, a simple yet effective framework for training retrievers directly from deep research agent interactions. LRAT progressively mines relevance signals from trajectories to construct high-quality query-document supervision, and then optimizes a dense retriever with utility-aware weighting. An overview of the framework is shown in \autoref{fig:method}.

\subsection{Mining Relevance Signals}
\label{sec:signal_mining}

Given a collection of valid agent trajectories $\{\mathcal{T}\}$ (Section~\ref{sec:env_setup}), our goal is to extract query-document supervision that reflects how agents actually consume and utilize retrieval results. Guided by the analysis in Section~\ref{sec:analysis}, we adopt a multi-stage, progressively refined relevance mining procedure. We start from coarse trajectory signals (browsing decisions) to obtain naive supervision, and then refine positives using post-browse reasoning traces to reduce noise.

\subsubsection{Naive Relevance Mining from Search-Browse Transitions}
\label{sec:naive_mining}

We first construct coarse supervision from the agent's \textbf{[Search]} $\rightarrow$ \textbf{[Browse]} transitions. Consider a search turn $t$ where the agent issues an intermediate query $q_t$ and receives a ranked top-$K$ candidate set
$\mathcal{D}_t = \{d_{t,i}\}_{i=1}^{K}$.
If the agent subsequently performs a [Browse] action on one of the candidates at the next turn, we view the browsed document $d_{t+1}$ as a \emph{naive positive} sample, since browsing is a necessary prerequisite for successful task completion (as analyzed in Section~\ref{sec:browse_signal}). 

For negatives, human click logs require careful debiasing due to position bias. In contrast, our analysis in Section~\ref{sec:negative_signal} shows that agent browsing decisions exhibit weak position dependence, suggesting that unbrowsed items within the retrieved candidate set are more likely to reflect explicit rejection rather than lack of exposure. Therefore, for each browsed document $d_{t+1}$, we treat all other candidates in the same retrieved set that are \emph{not browsed} as naive negatives:
\[
\mathcal{N}_t = \mathcal{D}_t \setminus \{d_{t+1}\}.
\]
This yields coarse training instances of the form $(q_t, d_{t+1}, \mathcal{N}_t)$.

\subsubsection{Reasoning-Aware Positive Filtering}
\label{sec:reasoning_filter}

Browsing actions are still imperfect proxies of relevance, because agents choose documents to browse based on coarse snippets and may later judge that a browsed document is unhelpful. Our analysis in Section~\ref{sec:reasoning_signal} shows that post-browse reasoning traces provide a reliable indicator of document utility, often explicitly stating whether the browsed content resolves the information gap.

Based on this insight, we refine naive positives with a \emph{reasoning-aware LLM-as-judge filter}. For each browsed document $d_{t+1}$, we collect the agent's immediate post-browse reasoning trace $r_{t+2}$ and apply an LLM-based verifier to determine whether the reasoning explicitly uses the document content to support progress on the task. Concretely, we use \textit{Qwen3-30B-A3B-Thinking-2507} as the judge and label $(q_t, d_{t+1})$ as \texttt{Relevant} or \texttt{Irrelevant} based on the reasoning trace $r_{t+2}$. 

This filtering step removes clear noise among browsed-but-unhelpful documents while preserving high-quality positives. In our empirical validation on BrowseComp-Plus evidence annotations, the verifier can retain \textbf{97.2\%} of ground-truth evidence documents, ensuring near-perfect recall of strong positives. Meanwhile, it retains \textbf{74.8\%} of browsed non-evidence documents, indicating that the filter removes obvious noise while still capturing agent-specific utility that may go beyond rigid dataset evidence labels.

\subsection{Intensity-Aware Training}
\label{sec:training}

We next describe how LRAT leverages the trajectory-derived supervision constructed in Section~\ref{sec:signal_mining} to train a retrieval model. Beyond identifying which documents are relevant, agent trajectories also reveal \emph{how strongly} a document contributes to task progress. LRAT explicitly incorporates this notion of relevance intensity into retriever optimization.

\subsubsection{Reasoning-Length Induced Relevance Intensity Estimation}
\label{sec:weight}

As analyzed in Section~\ref{sec:reasoning_signal}, post-browse reasoning provides a reliable indicator of document utility: longer reasoning chains following a browsing action are strongly correlated with higher document usefulness for the agent’s subsequent planning and problem solving. 
This phenomenon is analogous to classical human search, where dwell time has long been recognized as an effective proxy for relevance intensity~\cite{kelly2004display, kim2014modeling}. In both cases, increased cognitive effort reflects deeper engagement with the retrieved content.

\begin{wrapfigure}{r}{0.46\textwidth}
    \vspace{-0.8em}
    \centering
    \includegraphics[width=\linewidth]{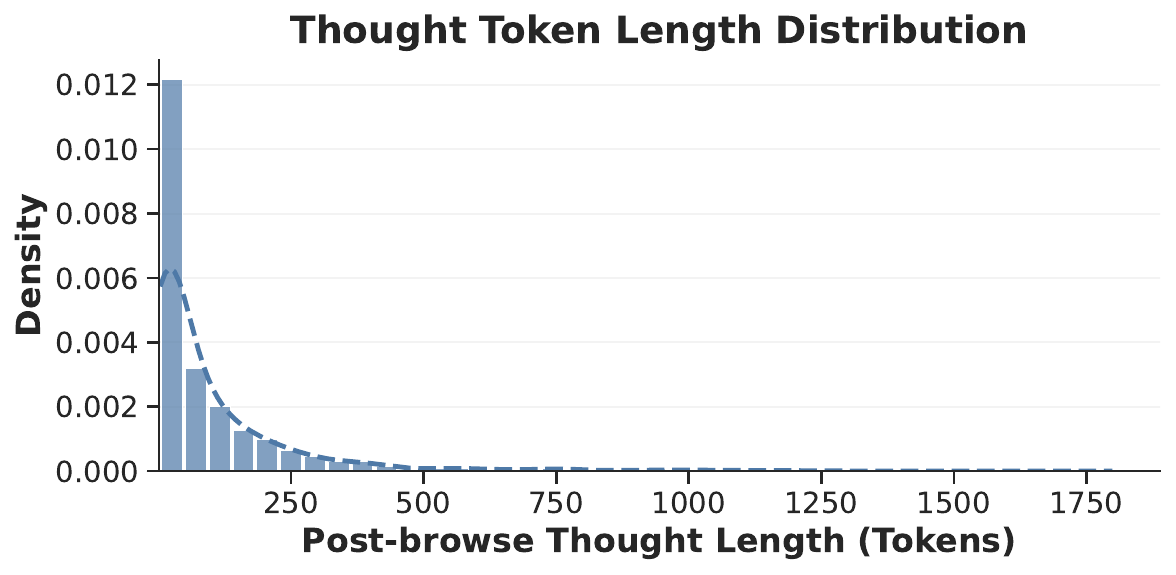}
    \vspace{-10pt}
    \caption{Distribution of thought token lengths after browsing actions, computed from agent trajectories in the same setting as \autoref{fig:traj_analysis}.}
    \label{fig:token_distribution}
    \vspace{-1.0em}
\end{wrapfigure}

Motivated by this analogy, we introduce a relevance intensity estimation scheme tailored to agent-derived feedback. Rather than treating all positive signals equally, we model relevance as a continuous quantity derived from the agent’s post-browse reasoning length. Inspired by the saturation principle underlying the time-aware click model~\cite{liu2016time}, we map reasoning length to a bounded utility score using an exponential saturation function, capturing the diminishing returns of increasingly long reasoning traces.

In the time-aware click model~\cite{liu2016time}, the marginal gain at dwell length $x$ follows an exponentially decaying function,
\begin{equation}
g(x) = \exp\left(-\frac{\ln 2}{\beta} x \right),
\end{equation}
where $\beta$ is the half-life of cumulative gain. Our analysis of post-browse reasoning lengths (Figure~\ref{fig:token_distribution}) shows a similar exponential decay in agent trajectories. Integrating this marginal gain from $0$ to the observed reasoning length $l$ yields the cumulative relevance utility:
\begin{equation}
u(l) = \int_0^l g(x)\, dx
      = \frac{\beta}{\ln 2} \left( 1 - \exp\left(-\frac{\ln 2}{\beta} l \right) \right),
\end{equation}
which naturally saturates for long reasoning traces. Normalizing across the dataset produces the relevance intensity weight used for training.

Formally, let $l = \text{Length}(r)$ denote the number of tokens in the reasoning trace immediately following the browsing of a document. We compute the relevance intensity weight as:
\begin{equation}
w = \frac{1}{\mu_{\text{raw}}} \left( 1 - \exp\left(-\frac{\ln 2 \cdot l}{\beta}\right) \right)
,
\label{eq:w}
\end{equation}

\noindent where $\beta$ is a length-scale parameter set to the median reasoning length across all trajectories, and $\mu_{\text{raw}}$ denotes the global mean of the unnormalized scores over the dataset. This normalization ensures $\mathbb{E}[w] \approx 1$, maintaining training stability while assigning higher importance to documents that trigger deeper agent reasoning and greater task progress. For simplicity, we omit the constant factor $\frac{\beta}{\ln 2}$, as the final weights are normalized across the dataset and the relative ranking of documents is preserved.

\subsubsection{Weighted Contrastive Learning}
\label{sec:loss}

We instantiate LRAT using a standard bi-encoder dense retriever, as commonly used in embedding models. Each query $q$ and document $d$ is independently encoded into vector representations $\mathbf{e}_q, \mathbf{e}_d \in \mathbb{R}^h$, where $h$ denotes the embedding dimension. The relevance score is then computed via a similarity function, such as dot product or cosine similarity: $s(q,d) = \mathrm{sim}(\mathbf{e}_q, \mathbf{e}_d)$.

To incorporate relevance intensity into optimization, we modify the standard InfoNCE loss \cite{gutmann2010noise} by introducing sample-wise weighting to train the dense retriever. For a mini-batch of size $N$, the weighted contrastive objective is defined as
\[
\mathcal{L} = - \frac{1}{N} \sum_{i=1}^{N} 
w_i \cdot 
\log \frac{\exp\left(s(q_i, d_i^+) / \tau\right)}
{\displaystyle \exp\left(s(q_i, d_i^+) / \tau\right) + 
\sum_{d^- \in \mathcal{N}_i} \exp\left(s(q_i, d^-) / \tau\right)},
\]
where $\tau$ is the temperature. $d_i^+$ denotes the positive document for query $q_i$, and $w_i$ is the reasoning-length–derived weight defined in Eq.~(\ref{eq:w}). The weight $w_i$ scales the gradient contribution of each instance, allowing documents associated with deeper reasoning to exert a stronger influence during training.

The negative set $\mathcal{N}_i$ is constructed from two complementary sources: (1) unbrowsed documents from the same retrieved candidate set, derived from agent trajectories as described in Section~\ref{sec:naive_mining}, and (2) in-batch negatives consisting of documents paired with other queries in the same mini-batch. This hybrid negative sampling strategy improves discriminative power, helping the retriever separate high-utility evidence from explicitly rejected candidates and unrelated documents, while avoiding representation collapse.
\section{Experiments}\label{sec:training}

In this section, we conduct extensive experiments to evaluate the effectiveness of our proposed LRAT framework in agentic search settings. 

\subsection{Experimental Setup}

\subsubsection{Benchmarks}
We evaluate LRAT on two benchmarks to assess both in-domain and out-of-domain generalization. For \textbf{in-domain evaluation}, we use \textbf{InfoSeek-Eval}~\cite{luo2025infoflow}, which consists of 300 multi-hop information-seeking queries that are strictly disjoint from all training data in Section~\ref{sec:env_data}. For \textbf{out-of-domain evaluation}, we additionally evaluate on \textbf{BrowseComp-Plus}~\cite{chenbrowsecomp}, a reproducible benchmark specifically designed for deep research agents. It contains 830 complex, human-authored questions that require multi-step reasoning and evidence aggregation. Following the official protocol, retrieval is performed over a corpus of 100,195 documents.

\definecolor{backblue}{RGB}{220, 230, 250}
\newcommand{\high}{\cellcolor{backblue}\textbf}

\begin{table*}[t!]
\centering
\renewcommand{\arraystretch}{0.8}
\setlength{\tabcolsep}{6pt}

\caption{
Results on task-optimized search agents and generalist agentic foundation models
with different retrievers on in-domain (ID) InfoSeek-Eval and out-of-domain (OOD) BrowseComp-Plus benchmarks.
Qwen3-Emb denotes Qwen3-Embedding-0.6B and E5-Large denotes Multilingual-E5-Large-Instruct.
Metrics include Success Rate (SR), Recall, and Average Step Count (Avg. Steps).
Best results within each agent backbone are highlighted in \textbf{bold}.
}
\vspace{-0.05cm}
\label{tab:main_results}

\resizebox{0.95\textwidth}{!}{%
\begin{tabular}{ll cc ccc}
\toprule
\multirow{2}{*}{\textbf{Agent Backbone}} & \multirow{2}{*}{\textbf{Retriever}} &
\multicolumn{2}{c}{\textbf{InfoSeek-Eval} (ID)} &
\multicolumn{3}{c}{\textbf{BrowseComp-Plus} (OOD)} \\
\cmidrule(lr){3-4} \cmidrule(lr){5-7}
 & & SR ($\uparrow$) & Avg. Steps ($\downarrow$) &
 SR ($\uparrow$) & Recall ($\uparrow$) & Avg. Steps ($\downarrow$) \\
\midrule

\rowcolor{sectionblue}
\multicolumn{7}{l}{\textsc{\textbf{I. Task-Optimized Search Agents}}} \\
\midrule

\multirow{4}{*}{\textbf{AgentCPM-Explore} (4B)}
 & Qwen3-Emb & 40.3 & 38.0 & 13.5 & 23.2 & 40.7 \\
 & \textbf{+ LRAT (Ours)} &
 \textbf{55.7 \textcolor{red}{($+38.2\%$)}} & \textbf{34.4} &
 \textbf{15.8 \textcolor{red}{($+17.0\%$)}} &
 \textbf{32.0 \textcolor{red}{($+37.9\%$)}} & \textbf{40.4} \\
\cmidrule(l{0.5em}r{0.5em}){2-7}
 & E5-Large & 47.3 & 38.9 & 15.9 & 26.5 & 40.7 \\
 & \textbf{+ LRAT (Ours)} &
 \textbf{49.7 \textcolor{red}{($+5.1\%$)}} & \textbf{35.5} &
 \textbf{15.9 ($+0.0\%$)} &
 \textbf{32.1 \textcolor{red}{($+21.1\%$)}} & \textbf{40.1} \\
\midrule

\multirow{4}{*}{\textbf{WebExplore} (8B)}
 & Qwen3-Emb & 52.0 & 24.1 & 21.0 & 47.7 & 40.7 \\
 & \textbf{+ LRAT (Ours)} &
 \textbf{68.7 \textcolor{red}{($+32.1\%$)}} & \textbf{19.0} &
 \textbf{27.2 \textcolor{red}{($+29.5\%$)}} &
 \textbf{55.9 \textcolor{red}{($+17.2\%$)}} & \textbf{38.7} \\
\cmidrule(l{0.5em}r{0.5em}){2-7}
 & E5-Large & 60.0 & 23.8 & 25.4 & 50.4 & 40.1 \\
 & \textbf{+ LRAT (Ours)} &
 \textbf{63.3 \textcolor{red}{($+5.5\%$)}} & \textbf{20.2} &
 \textbf{29.0 \textcolor{red}{($+14.2\%$)}} &
 \textbf{56.1 \textcolor{red}{($+11.3\%$)}} & \textbf{39.1} \\
\midrule

\multirow{4}{*}{\textbf{Tongyi-DeepResearch} (30B)}
 & Qwen3-Emb & 52.7 & 26.7 & 17.8 & 49.2 & 42.9 \\
 & \textbf{+ LRAT (Ours)} &
 \textbf{68.0 \textcolor{red}{($+29.0\%$)}} & \textbf{20.7} &
 \textbf{23.7 \textcolor{red}{($+33.1\%$)}} &
 \textbf{60.7 \textcolor{red}{($+23.4\%$)}} & \textbf{41.0} \\
\cmidrule(l{0.5em}r{0.5em}){2-7}
 & E5-Large & 56.7 & 25.1 & 20.7 & 54.8 & 42.4 \\
 & \textbf{+ LRAT (Ours)} &
 \textbf{68.0 \textcolor{red}{($+19.9\%$)}} & \textbf{21.5} &
 \textbf{23.9 \textcolor{red}{($+15.5\%$)}} &
 \textbf{61.8 \textcolor{red}{($+12.8\%$)}} & \textbf{41.4} \\
\midrule

\rowcolor{sectionblue}
\multicolumn{7}{l}{\textsc{\textbf{II. Generalist Agentic Foundation Models}}} \\
\midrule

\multirow{4}{*}{\textbf{GPT-OSS} (120B)}
 & Qwen3-Emb & 40.0 & 34.9 & 9.0 & 43.7 & 45.4 \\
 & \textbf{+ LRAT (Ours)} &
 \textbf{47.0 \textcolor{red}{($+17.5\%$)}} & \textbf{30.5} &
 \textbf{12.1 \textcolor{red}{($+34.4\%$)}} &
 \textbf{56.4 \textcolor{red}{($+29.1\%$)}} & \textbf{45.2} \\
\cmidrule(l{0.5em}r{0.5em}){2-7}
 & E5-Large & 41.7 & 33.9 & 10.8 & 50.1 & 44.8 \\
 & \textbf{+ LRAT (Ours)} &
 \textbf{50.7 \textcolor{red}{($+21.6\%$)}} & \textbf{29.7} &
 \textbf{13.1 \textcolor{red}{($+21.3\%$)}} &
 \textbf{56.0 \textcolor{red}{($+11.8\%$)}} & \textbf{44.6} \\
\midrule

\multirow{4}{*}{\textbf{MiniMax-M2.1} (229B)}
 & Qwen3-Emb & 58.7 & 21.4 & 38.2 & 57.2 & 30.8 \\
 & \textbf{+ LRAT (Ours)} &
 \textbf{78.3 \textcolor{red}{($+33.4\%$)}} & \textbf{14.7} &
 \textbf{48.3 \textcolor{red}{($+26.4\%$)}} &
 \textbf{69.2 \textcolor{red}{($+21.0\%$)}} & \textbf{28.3} \\
\cmidrule(l{0.5em}r{0.5em}){2-7}
 & E5-Large & 64.0 & 18.9 & 46.4 & 64.9 & 29.1 \\
 & \textbf{+ LRAT (Ours)} &
 \textbf{75.0 \textcolor{red}{($+17.2\%$)}} & \textbf{14.8} &
 \textbf{48.7 \textcolor{red}{($+5.0\%$)}} &
 \textbf{69.7 \textcolor{red}{($+7.4\%$)}} & \textbf{28.9} \\
\midrule

\multirow{4}{*}{\textbf{GLM-4.7} (358B)}
 & Qwen3-Emb & 67.7 & 27.5 & 43.9 & 66.6 & 45.5 \\
 & \textbf{+ LRAT (Ours)} &
 \textbf{82.0 \textcolor{red}{($+21.1\%$)}} & \textbf{18.5} &
 \textbf{54.6 \textcolor{red}{($+24.4\%$)}} &
 \textbf{77.8 \textcolor{red}{($+16.8\%$)}} & \textbf{44.6} \\
\cmidrule(l{0.5em}r{0.5em}){2-7}
 & E5-Large & 73.7 & 24.2 & 46.4 & 68.7 & \textbf{44.6} \\
 & \textbf{+ LRAT (Ours)} &
 \textbf{81.7 \textcolor{red}{($+10.9\%$)}} & \textbf{19.5} &
 \textbf{50.6 \textcolor{red}{($+9.1\%$)}} &
 \textbf{76.3 \textcolor{red}{($+11.1\%$)}} & 44.8 \\
\bottomrule
\end{tabular}%
}
\vspace{-0.1cm}
\end{table*}

\subsubsection{Backbones}
For the retrieval backbone, we consider two representative and widely adopted dense retrievers with complementary architectures:
\textbf{Multilingual-E5-Large-Instruct}~\cite{wang2024multilingual}, an encoder-based model, and
\textbf{Qwen3-Embedding-0.6B}~\cite{qwen3embedding}, a decoder-based embedding model.

To evaluate downstream impact in realistic agentic settings, we integrate these retrievers into multiple open-source search agents to ensure reproducibility. These include \textbf{task-optimized search agents} (\textbf{AgentCPM-Explore}-4B~\cite{AgentCPMExplore2026}, \textbf{WebExplore}-8B~\cite{liu2025webexplorer}, \textbf{Tongyi-DeepResearch}-30B~\cite{team2025tongyi}) and \textbf{generalist agentic foundation models} (\textbf{GPT-OSS}-120B~\cite{openai2025gptoss120bgptoss20bmodel}, \textbf{MiniMax-M2.1}-229B~\cite{minimax2025m21}, \textbf{GLM-4.7}-358B~\cite{5team2025glm45agenticreasoningcoding}).  
This setup spans diverse retriever architectures and agent scales ranging from 4B to 358B parameters, enabling a comprehensive evaluation of LRAT across different agentic search systems.

\subsubsection{Evaluation Metrics} 
Following the standard evaluation protocol of BrowseComp-Plus~\cite{chenbrowsecomp}, we adopt a multi-dimensional evaluation setup that assesses task outcome, execution efficiency, and retrieval quality.
The primary metric is \textbf{Success Rate (SR)}, which is assessed by an automated LLM judge based on \textit{Qwen3-30B-A3B-Thinking-2507} to verify answer correctness.
To measure execution efficiency, we report the \textbf{Average Step Count}, where fewer steps indicate more direct information acquisition enabled by the retriever.
To directly measure retrieval quality, we additionally report \textbf{Evidence Recall} on BrowseComp-Plus, defined as the proportion of tasks in which the annotated evidence document is successfully retrieved during the agent's execution.
For InfoSeek-Eval, we report success rate and average step count only, as it does not provide the trace-level annotations required to compute evidence recall.

\subsubsection{Implementation Details}
We train the retriever with the FlagEmbedding framework~\cite{flagopen2023flagembedding}. The model is fine-tuned for 2 epochs, with a batch size of 32, a learning rate of $1\mathrm{e}{-6}$, and a maximum input length of 512 tokens. For InfoNCE loss, we use a group size of 10 and a temperature of 0.02. On the agent side, for reproducibility, we fix the random seed to 2025. Generation parameters are set to a temperature of 0.85, $\text{top}_p=0.95$, and a presence penalty of 1.1. Retrieval configuration during execution is kept consistent with the training setup. Trajectories are limited to 50 turns per query during evaluation due to computational constraints. 

\subsection{Overall Performance}

\autoref{tab:main_results} reports the main results on both in-domain and out-of-domain benchmarks. Overall, LRAT yields consistent and substantial improvements across all retriever backbones and agent backbones, leading to higher task success, stronger evidence retrieval, and more efficient agent execution. Specifically, we have the following observations:

\textbf{Improved Evidence Retrieval.}
On BrowseComp-Plus, LRAT significantly improves the retriever’s ability to retrieve annotated evidence documents, as reflected by the recall metric. For both Qwen3-Embed and E5-Large, training with LRAT consistently increases evidence recall across all agents, with relative gains ranging from 7\% to over 37\%. This demonstrates that supervision derived from agent trajectories effectively enhances retrieval quality, enabling retrievers to better align with the information needs issued by agents.

\textbf{End-to-end Gains in Task Success.}
Stronger retrieval quality directly translates into improved end-to-end performance. Across both in-domain and out-of-domain settings, agents equipped with LRAT-trained retrievers achieve substantially higher success rates than their baseline counterparts. These gains are consistent across task-optimized search agents and generalist agentic foundation models, indicating that LRAT generalizes well across agent architectures and parameter scales. Notably, improvements persist even for very large agents (e.g., 120B-358B), suggesting that retrieval quality remains a critical bottleneck despite strong agent capabilities.

\textbf{More Efficient Agent Execution.}
In addition to higher success rates, LRAT consistently reduces the average number of interaction steps required to solve a task. This effect is particularly pronounced on InfoSeek-Eval, where average step counts are reduced by up to \(\sim 30\%\). The reduction in steps indicates that LRAT-trained retrievers provide more precise and useful evidence at each search step, allowing agents to satisfy their information needs with fewer exploratory interactions. As a result, agents achieve better performance while incurring fewer search and browse actions.

\subsection{Ablation Study}
\label{sec:ablation}

\begin{figure}[t]
    \centering
    \includegraphics[width=0.75\linewidth]{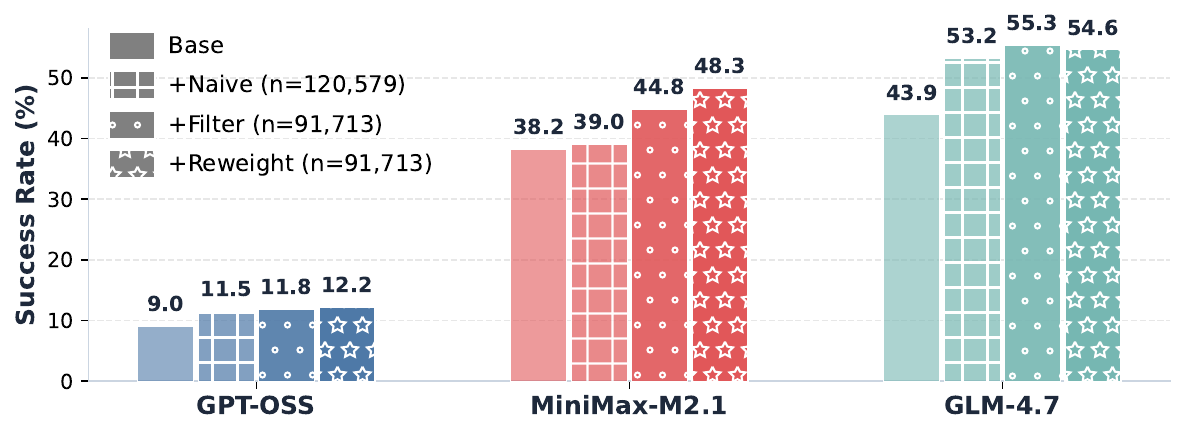}
    \vspace{-0.35cm}
    \caption{
    Ablation study: components are incrementally added. Numbers $n$ in parentheses show the amount of training data used for each variant of LRAT.
    }
    \vspace{-0.2cm}
    \label{fig:ablation_study}
\end{figure}

We conduct an ablation study on BrowseComp-Plus using Qwen3-Embedding-0.6B as the retriever, with components of LRAT added incrementally. Unless otherwise specified, subsequent experiments use the same setting. The results in \autoref{fig:ablation_study} highlight the significance of each design:

\textbf{``+Naive''} refers to a variant that treats only the documents browsed by the agent as positive samples, while regarding all other documents as negatives.
This strategy yields substantial performance gains, suggesting the absence of strong position bias in the agent's browsing process and confirming that unbrowsed documents can serve as reliable negative signals.

\textbf{``+Filter''} further introduces LLM-based filtering over the browsed documents, removing false positive documents that are accessed but do not meaningfully contribute to the agent’s subsequent reasoning.
The resulting improvement indicates that post-browse reasoning traces are important indicators of document usefulness, and leveraging them helps refine the quality of supervision signals.

\textbf{``+Reweight''} incorporates relevance intensity estimation by using the reasoning length of the agent as a proxy for importance.
The resulting performance gains highlight the necessity of accounting for the heterogeneous contributions of different documents and empirically validate the effectiveness of LRAT in leveraging reasoning-aware supervision.

Overall, the ablation results empirically validate the progressive design of LRAT, showing that trajectory-derived supervision becomes increasingly effective as richer agent signals are mined and incorporated into retriever training.

\subsection{Scalability and Robustness Analysis}
\label{sec:scaling}

\begin{figure}[h]
    \centering
    \includegraphics[width=0.8\linewidth]{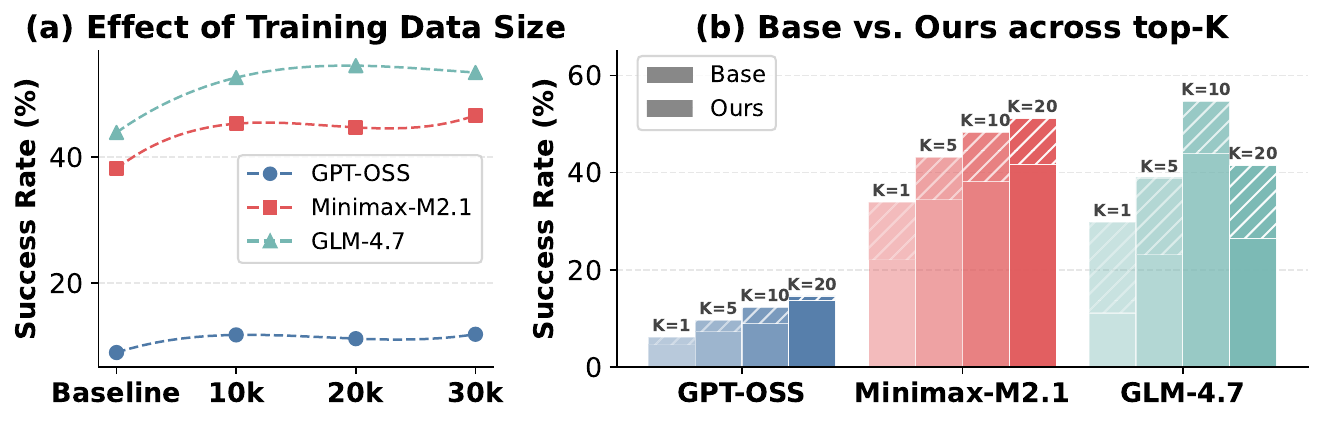}
    \vspace{-0.3cm}
    \caption{Agent performance with varying training data sizes and retrieval top-$K$ settings.}
    \vspace{-0.15cm}
    \label{fig:scale}
\end{figure}

To assess whether LRAT remains effective at realistic scales, 
we analyze its behavior under increasing training data and varying top-$K$ retrieval budgets. 

\textbf{Training-time Scalability.} We investigate whether retrievers trained with LRAT can consistently benefit from additional agent interaction data.
Specifically, we collect 30K agent trajectories on InfoSeekQA using the Tongyi-DeepResearch agent with BM25, and train retrievers with progressively larger subsets of the collected data.
As shown in \autoref{fig:scale}(a), agent success rates generally improve as the training dataset grows across all evaluated models, indicating that LRAT can effectively exploit larger volumes of agent trajectories and does not suffer from early performance saturation.

\textbf{Inference-time Robustness.} We further evaluate agent performance under different top-$K$ retrieval settings by comparing the base retriever with the LRAT-enhanced retriever. 
As illustrated in \autoref{fig:scale}(b), increasing $K$ does not always lead to monotonic performance improvements: while moderate retrieval budgets are beneficial, overly large $K$ can degrade performance (e.g., GLM-4.7), likely due to increased noise and limited effective context capacity.
Despite this, LRAT consistently outperforms the base retriever across all evaluated top-$K$ values, demonstrating its robustness to varying retrieval budgets and its effectiveness under both information-scarce and noise-heavy retrieval conditions.

\subsection{Data Flywheel Simulation and Analysis}\label{sec:flywheel}

\begin{table}[t]
\caption{Trajectory correctness ablation. Both correct and incorrect trajectories use 10K examples each.}
\vspace{-5pt}
\label{tab:trajectory_ablation}
\centering
\resizebox{0.8\textwidth}{!}{%
\begin{tabular}{l c c c}
\toprule
\textbf{Training Data} & \textbf{GPT-OSS} & \textbf{MiniMax-M2.1} & \textbf{GLM-4.7} \\
\midrule
Base (\wo LRAT) 
 & 9.0 & 38.2 & 43.9 \\
LRAT (\w Incorrect Trajectories) 
 & 10.7 \textcolor{red}{($+18.9\%$)} 
 & 43.6 \textcolor{red}{($+14.1\%$)} 
 & 50.6 \textcolor{red}{($+15.3\%$)} \\
LRAT (\w Correct Trajectories) 
 & 11.8 \textcolor{red}{($+31.1\%$)} 
 & 45.3 \textcolor{red}{($+18.6\%$)} 
 & 52.6 \textcolor{red}{($+19.8\%$)} \\
\bottomrule
\end{tabular}%
\vspace{-0.2cm}
}
\end{table}

Traditionally, human click logs are continuously exploited to iteratively improve retrievers, creating a self-sustaining data flywheel. We investigate whether a similar mechanism can arise under the LRAT framework in agentic search settings. Unlike benchmark evaluation, real user queries are open-ended, and agent trajectories are not always fully correct, raising the question of whether imperfect trajectories can still provide useful supervision. Our preliminary analysis suggests that this is indeed the case.
As shown in Table~\ref{tab:trajectory_ablation}, retrievers trained with both correct and incorrect trajectories consistently outperform the base retriever, although incorrect trajectories yield smaller gains.
This indicates that even when an agent fails to produce the correct final answer, its intermediate interactions with the retriever still reflect meaningful judgments about document utility. Thus, we can include all collected trajectories when correctness labels are unavailable or unreliable.

\begin{figure}[t]
    \centering
    \setlength{\abovecaptionskip}{4pt}
    \setlength{\belowcaptionskip}{-2pt}
    \begin{subfigure}[t]{0.485\linewidth}
        \centering
        \includegraphics[width=\linewidth]{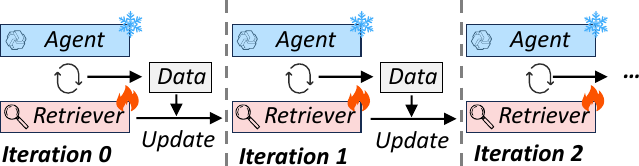}
        \caption{Illustration of data flywheel simulation setting.}
        \label{fig:flywheel_setting}
    \end{subfigure}
    \hfill
    \begin{subfigure}[t]{0.485\linewidth}
        \centering
        \includegraphics[width=\linewidth]{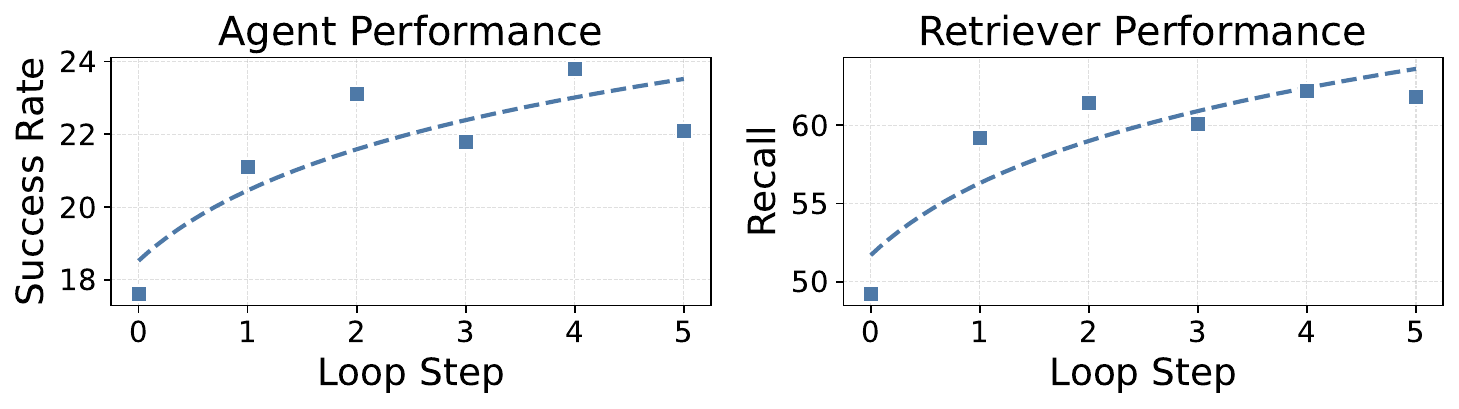}
        \caption{Performance of the data flywheel simulation.}
        \label{fig:data_flywheel}
    \end{subfigure}
\vspace{0.2em}
    \caption{Data flywheel simulation in the single-column layout. Left: the iterative update setting used to collect and refresh retriever supervision. Right: the resulting performance trend, showing steady gains in both success rate and evidence recall across iterations.}
    \vspace{-0.25cm}
\end{figure}

The simulated data flywheel is illustrated in \autoref{fig:flywheel_setting}, where the retriever undergoes iterative updates through continuous agent interactions.
At each step, the retriever collects trajectories from agent interactions and is updated before the next step, mimicking a realistic streaming environment.
To ensure consistency with our main experiments while controlling computational cost, we adopt the Tongyi-DeepResearch agent and sample 10K queries from InfoSeekQA at each step.
Evaluation is conducted using the same agent, reflecting a realistic deployment scenario with repeated retriever–agent interactions.
The results in \autoref{fig:data_flywheel} show steady improvements in both agent success rate and retriever recall across iterations, demonstrating that our method can reliably support iterative retriever updates and sustain a positive data flywheel. 
Moreover, performance is maintained or even improved in this streaming setting, highlighting the practical value of agent-based trajectory supervision for real-world retrieval systems.

\section{Conclusion}
This paper identifies a fundamental misalignment between human-centric retrieval training and the needs of agentic search, and formalizes \emph{learning to retrieve from agent trajectories} as a new retrieval paradigm.
Empirical analysis of deep research agent trajectories reveals that browsing behavior can serve as a reliable indicator of document utility, unbrowsed results can provide trustworthy negative signals, and post-browse reasoning traces can capture the intensity of document relevance during multi-step problem solving. A retrieval framework called LRAT is proposed to convert agent trajectories into supervision signals to train agent-aligned retrievers. Experiments on both in-domain and out-of-domain deep research benchmarks demonstrate the effectiveness of LRAT across diverse agent and retriever backbones. Empirical analysis also demonstrates that agent trajectories can support iterative retriever improvement, indicating the potential for a sustainable data flywheel driven by agent interactions. These findings highlight agent trajectories as a practical and scalable supervision source, and suggest a promising direction for advancing retrieval systems in the era of agentic search.

\bibliography{main}

\end{document}